\documentclass[twocolumn]{IEEEtran}
\usepackage[T1]{fontenc}
\usepackage{color}
\usepackage{float}
\usepackage{mathrsfs}
\usepackage{amsmath}
\usepackage{amssymb}
\usepackage{graphicx}
\usepackage{cite}
\usepackage{makecell}
\usepackage{fancyhdr}
\usepackage{diagbox}
\usepackage{array}
\usepackage[raggedrightboxes]{ragged2e}
\usepackage{color}
\usepackage[justification=justified,singlelinecheck=false]{caption}

\usepackage[unicode=true,
 bookmarks=true,bookmarksnumbered=true,bookmarksopen=true,bookmarksopenlevel=1,
 breaklinks=false,pdfborder={0 0 0},pdfborderstyle={},backref=false,colorlinks=false]
 {hyperref}
\hypersetup{pdftitle={Your Title},
 pdfauthor={Your Name},
 pdfpagelayout=OneColumn, pdfnewwindow=true, pdfstartview=XYZ, plainpages=false}

\makeatletter

\floatstyle{ruled}
\newfloat{algorithm}{tbp}{loa}
\providecommand{\algorithmname}{Algorithm}
\floatname{algorithm}{\protect\algorithmname}

\usepackage[caption=false,font=footnotesize]{subfig}
\usepackage{algorithm}
\usepackage{algorithmic}
\usepackage{subcaption}  
\usepackage{multirow} 
\usepackage{amsmath} 
\usepackage{xcolor}

\usepackage{makecell}

\allowdisplaybreaks[4]

\ifCLASSOPTIONcompsoc
\usepackage[caption=false,font=normalsize,labelfont=sf,textfont=sf]{subfig}
\else
\usepackage[caption=false,font=footnotesize]{subfig}
\fi

\usepackage{bm}
\usepackage{algorithmic}
\usepackage{algorithm}
\usepackage{graphicx}
\IEEEoverridecommandlockouts

\usepackage{lettrine}

\usepackage{geometry}
\@ifundefined{showcaptionsetup}{}{%
 \PassOptionsToPackage{caption=false}{subfig}}
\usepackage{subfig}

\makeatother

\usepackage{fancyhdr}
\begin{document}
\setlength{\parskip}{0em}
\title{\textcolor{black}{Hierarchical Low-Altitude Wireless  Network Empowered Air Traffic Management}}


\author{\IEEEauthorblockN{Ziye Jia, Jia He,  Yuanhao Cui, Qiuming Zhu, Ligang Yuan, Fuhui Zhou, Qihui Wu, \IEEEmembership{Fellow,~IEEE,} \\ Dusit Niyato, \IEEEmembership{Fellow,~IEEE,} and Zhu Han, \IEEEmembership{Fellow,~IEEE}}
\thanks{\justifying

Ziye Jia, Jia He, Qiuming Zhu, Fuhui Zhou and Qihui Wu are with the College of Electronic and Information Engineering, Nanjing University of Aeronautics and Astronautics, Nanjing 211106, China, (e-mail: jiaziye@nuaa.edu.cn, 071940128hejia@nuaa.edu.cn, zhuqiuming@nuaa.edu.cn, zhoufuhui@nuaa.edu.cn, wuqihui@nuaa.edu.cn). 

Yuanhao Cui is with the School of Information and Communication Engineering, Beijing University of Posts and Telecommunications, Beijing 100876, China (e-mail: cuiyuanhao@bupt.edu.cn).

Ligang Yuan is with the College of Civil Aviation, Nanjing University of Aeronautics and Astronautics, Nanjing 211106, China (e-mail: yuanligang@nuaa.edu.cn).

Dusit Niyato is with the College of Computing and Data Science, Nanyang Technological University, Singapore 639798 (e-mail:
dniyato@ntu.edu.sg).

Zhu Han is with the University of Houston, TX 77004, USA (e-mail: zhan2@uh.edu), and also with the Department of Computer Science and Engineering, Kyung Hee University, Seoul, 446-701, South Korea.}}

\maketitle
\pagestyle{headings} 
\rhead{\thepage}
\renewcommand{\headrulewidth}{0pt}
\begin{abstract}
As the increasing development of low-altitude aircrafts, the rational design of low-altitude  networks directly impacts the aerial safety and resource utilization. 
To address the challenges of  environmental complexity and aircraft diversity in the  traffic management, we propose a hierarchical low-altitude  wireless network (HLWN) framework. 
Empowered by the three-dimensional spatial discretization and integrated wireless monitoring mechanisms in HLWN, 
we design low-altitude air corridors to  guarantee safe operation and optimization.
Besides, we develop the multi-dimensional flight risk assessment through conflict detection and probabilistic collision analysis, facilitating dynamic collision avoidance for heterogeneous aircrafts. 
Finally, the open issues and future directions are investigated  to provide insights into  HLAN development.

\end{abstract}


\section{Introduction}\label{s1}
\IEEEPARstart{T}{he} low-altitude space is emerging as a significant economic driver that primarily focuses on the airspace below 3,000 meters above the ground level \cite{ZHOU2025145050}.  It can be supported by the 5th generation advanced (5G-A) technologies to create an integrated air-ground system,  enabling applications such as air mobility, intelligent logistics, emergency services, and cultural tourism \cite{Economy}. 
Unmanned aerial vehicles (UAVs) and electric vertical take-off and landing (eVTOL) aircrafts are widely applied due to their respective advantages in maneuverability \cite{Multi-MEC}. 
The explosion number of heterogenous aircrafts  urge advanced air traffic management technologies to unlock the full potentials of the low-altitude airspace.

However, the mixed structure,  meteorological conditions, and various aircraft operations have all posed  challenges to the air traffic management in the low-altitude space, leading to insufficient space utilization and high collision risks \cite{zhangyifan_CJA}.
To address these challenges in high-density aerial mobility scenarios, the layered airspace structure, and  fine-grained grid based corridor design should be well considered \cite{Sensors2021_corridor, PIMRC_2023_Corridor_ground}.
In addition, the design of integrated low-altitude wireless surveillance mechanisms is essential to guarantee the real-time situational awareness of structured corridors, and achieve safe and efficient operations in dense airspaces  \cite{TWC_Corridor_cellular}. 
The realization of low-altitude wireless network will also benefit to the air traffic conflict management, but the existing conflict detection methodologies are inadequate for the multi-target and high-dynamic environments, and the trajectory optimization lacks multidimensional coordination. 
In summary, the major challenges in LAIN include the airspace complexity, collaborative monitoring mechanisms, and aerial risk management, detailed as follows.

\begin{figure*}[!t]
    \centering
    \includegraphics[width=11.5cm]{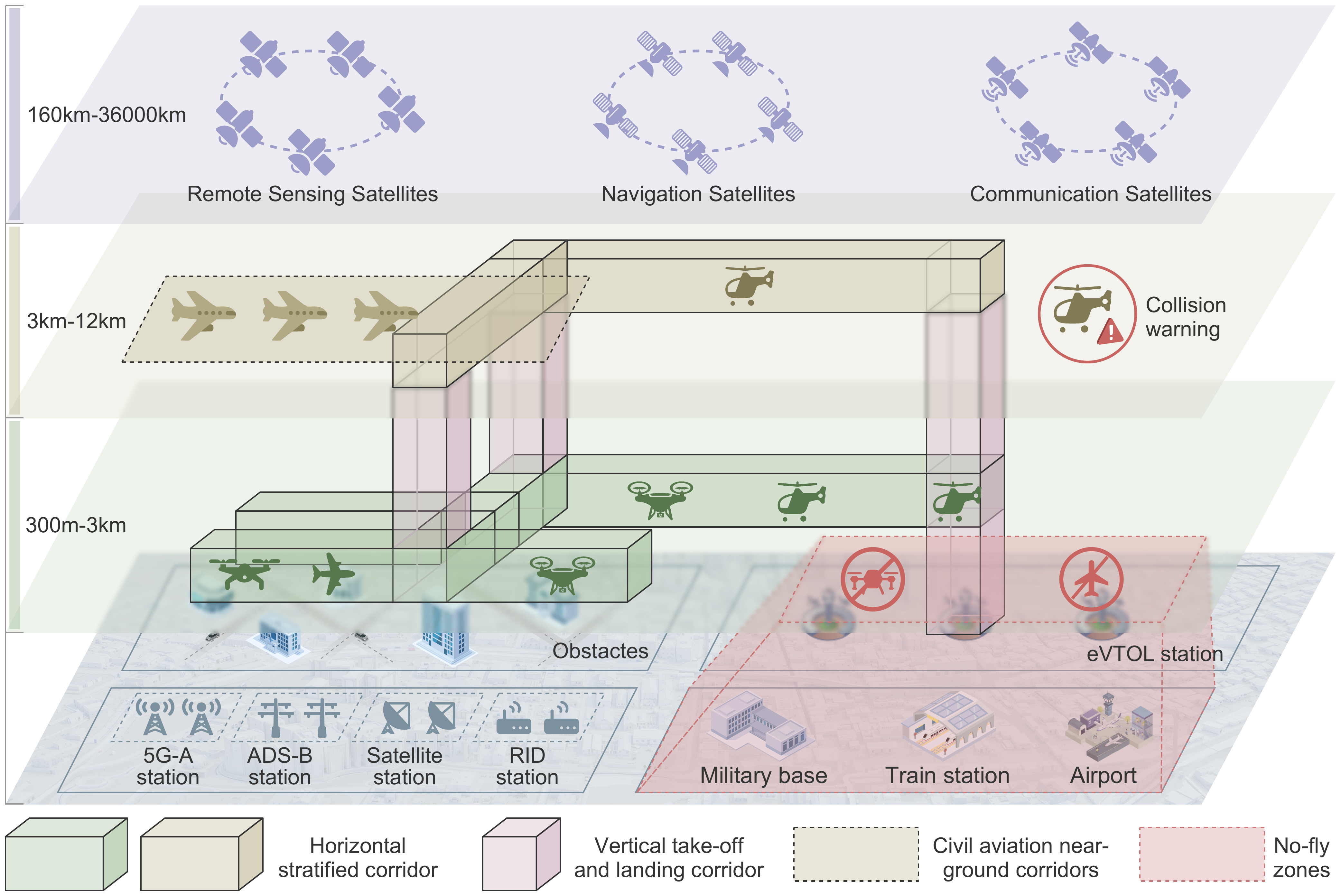}
    \caption{An illustration of HLWN framework, composed of multiple  layers, crossed corridors, heterogeneous aircrafts, supported by satellite services. The ground layer includes multiple wireless stations, obstacles, sensitive areas, and landing fields.} 
    \label{f1}
\end{figure*}

\begin{itemize}    
    \item \textbf{Airspace Complexity}: 
    The low-altitude airspace is characterized by a wide variety of aircraft types, heavy traffic density, and  complex electromagnetic interference environment \cite{zhuqiuming_TWC}, so it cannot directly leverage the traditional air traffic control systems. 
    In addition, the performance gaps in speed, maneuverability and communication among fixed-wing, rotary-wing, and hybrid aircrafts pose further challenges to the coordination models. 
    Consequently, a hierarchical air traffic management architecture is significant to deal with these complex flight characteristics. 
    
    \item \textbf{Collaborative Monitoring}: 
    The integration of multi-source data such as  aircraft state, meteorology, communication quality suffers from fragmented data standards and protocols, hindering the real-time coordination in the monitoring system. 
    The low airspace may simultaneously accommodate hundreds of aerial aircrafts, and the traditional servers often struggle to support the real-time processing of massive datasets. 
    Therefore, it is necessary to develop lightweight and distributed wireless monitoring model for the collaborative  surveillance.
    
    \item \textbf{Aerial Risk}: 
    The complexity and high dynamic in low-altitude airspace inherently increase the risk of collisions.
    It  will be particularly serious within aerial corridors, since the fine-grained grids of corridors further aggravate the complexity and danger of mid-air conflicts.  
    To effectively mitigate these hazards and prevent collisions, it is critical to  establish coordinated collision avoidance mechanisms among corridors, supported by the implementation of dynamic separation management within them. 
\end{itemize} 

In this work, we present the hierarchical low-altitude wireless network  (HLWN) framework, which is supported by the fine-grained airspace grid model and corridor design. 
Then, we propose the collaborative monitoring of multi-device surveillance system and multi-dimensional comprehensive evaluation mechanism. 
Through the airspace stratification and multi-modal wireless monitoring, the safe operation of heterogeneous aircrafts can also improve the utilization of low-altitude resources. 
Moreover, we develop the dynamic trajectory optimization model that reduces collision risks through adaptive trajectory planning and real-time hazard assessment.
The open issues and directions are also discussed.

\section{Framework of Hierarchical Low-altitude Wireless Network}\label{s2}  

As shown in Fig. \ref{f1},  based on the multi-layer airspace division, the HLWN framework clearly defines the airspace boundaries and restricted zones, establishes tiered functional hierarchies through differentiated airspace classifications. 
Also, the hierarchical corridor architecture is designed to optimize the airspace utilization. 
Furthermore, the application of heterogeneous aircrafts relies on collaborative monitoring technologies to achieve precise management and control. 
As a result, leveraging the communication, navigation, sensing and surveillance technology, HLWN can implement real-time flight monitoring, collision prediction, and dynamic aircraft scheduling. 
These features guarantee secure, orderly, and scalable air traffic management in low-altitude airspaces.    
Specifically, the hierarchical architecture of HLWN includes the airspace structure for traffic management, and wireless infrastructure for monitoring.

\begin{figure*}[!t]
    \centering
    \includegraphics[width=14.8cm]{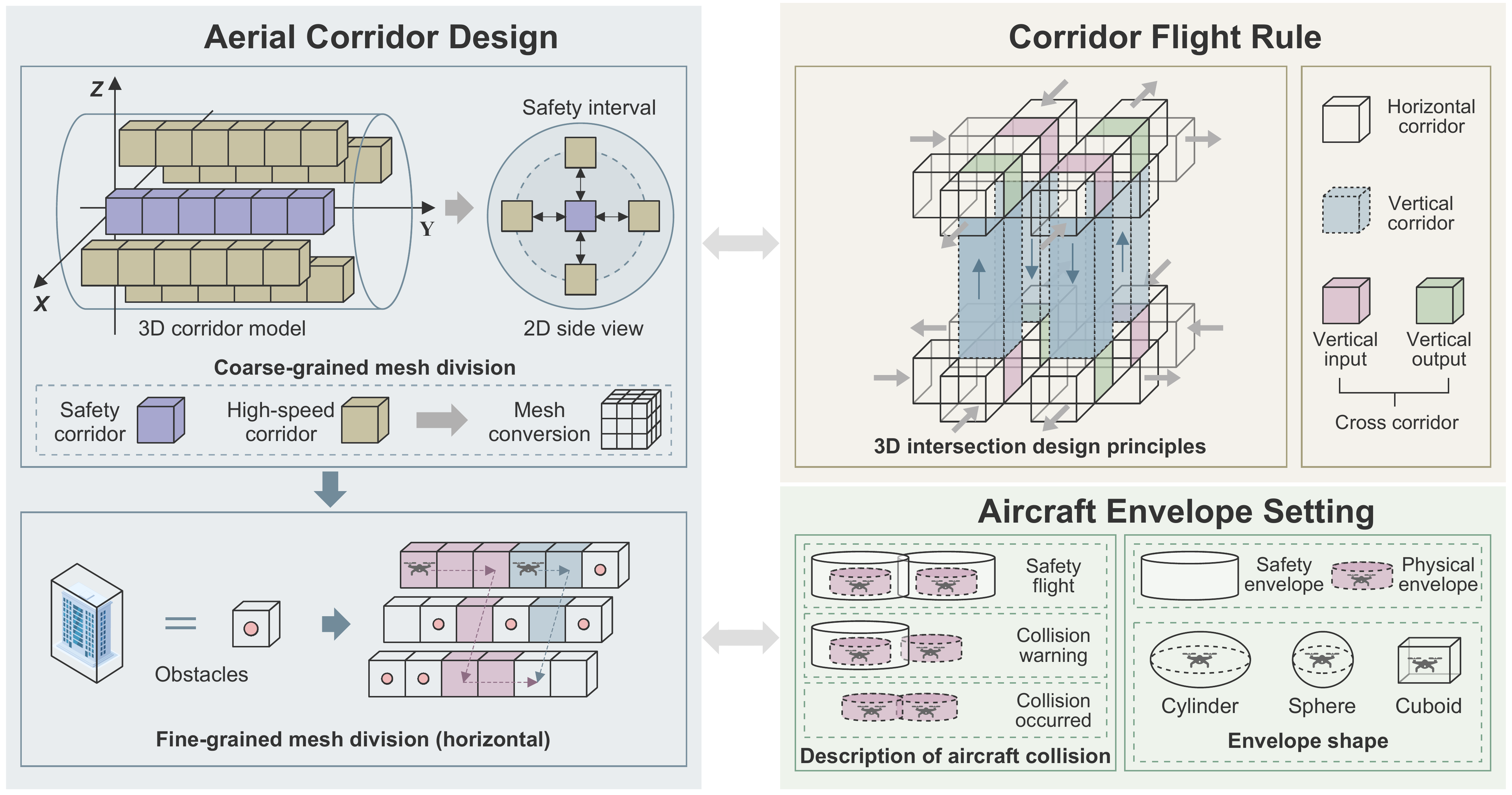}
    \caption{The grid segmentation based corridor design, including the vertical coarse-grained mesh division, and horizontal fine-grained mesh devision. The corridor flight rule and the aircraft envelope are also illustrated for the aerial safety management.}
    \label{f2}
\end{figure*}

\subsubsection{Airspace Structure}
The HLWN is composed of multiple layers, where the satellite layers provide auxiliary services for the aerial layers, and the terrestrial layer is mainly classified into multiple wireless ground stations for monitoring and communication, obstacles, sensitive areas, and landing fields. 
The no-fly zone boundaries are designed by geofences to limit the flight ranges for aerial safety.
The vertical corridors connect elevated vertiports  and support short-haul transit operations for the eVTOL aircrafts, and the stratified horizontal corridor separates the trajectories of unmanned aircraft and manned aircraft by altitudes. 
This  can help optimize low-altitude air traffics, by creating obstacle-free corridors, and incorporating the privacy-protected altitude rules (e.g., the aircrafts should maintain a minimum flight altitude to  avoid collisions with ground objects) and restricted zone (e.g., the airspace within 5km radius of an international airport). 
In particular, according to Chinese civil aviation regulations, the civil aviation aerial corridors above 300 meters can serve as buffer zones, ensuring the safe integrations with  traditional air traffics. 

\subsubsection{Wireless Infrastructure}
The wireless infrastructure establishes the integrated monitoring coordination based on the global navigation satellite system (GNSS), 5G-A, automatic dependent surveillance broadcast (ADS-B), and remote identification (RID).
In particular, GNSS provides global, high-precision positioning across all operational zones, while 5G-A networks deliver ultra-low latency, high-bandwidth data transmission \cite{yuanweijie_LAWN}, enabling real-time exchange of situational data critical for low-altitude traffic operations.
ADS-B facilitates the real-time broadcasting of aircraft location updates. 
RID provides rigorous identity verification processes for aircrafts.
These integrated components collectively synthesize dynamic airspace situational awareness and support the  multi-modal surveillance.  

In brief, the airspace structure establishes a spatial ordering of physical space through grid-based corridors, laying the foundations for operational activities. 
The wireless infrastructure illustrates multi-modal collaborative monitoring to dynamically generate the airspace situation awarenesses and support timely decision-making. 

\section{Low-Altitude Aerial Corridor Design}\label{s3}   
\subsection{Aerial Corridor Model}
The low-altitude airspace may be affected by ground human activities, uncoordinated aircraft operations, and the civil aviation aircrafts, which is a multi-layered hazard network, increasing the risks of collisions and airspace congestions. 
To address these challenges, HLWN can implement based on the well-designed 3D grid model \cite{TAES_2024_UAS_corridor} combined with tiered aerial safety management strategies. 
By dividing the 3D airspace into hierarchical aerial corridors and adaptive zones, HLWN allows more aircrafts to operate safely within a constrained safety space, thereby enhancing the airspace capacities.
Moreover, similar with the ground transportation relying  on traffic lights and lane markings, the aerial corridor model can establish  an ordered airspace traffic flow for low-altitude aircrafts by defining clear operational rules. 

As shown in Fig. \ref{f2}, the multi-functional hierarchical model includes four types of specialized corridors: horizontal corridors for aircraft movement, vertical corridors for altitude-layer management,  segregated corridors for high-speed traffics, and safety corridors for emergency transportations \cite{impact}. 
Specifically, the airspace grid employs a strategic division approach tailored to operational requirements at different altitudes \cite{Drone_Delivery}. 
The horizontal corridors facilitate general planed air traffics, while the vertical corridors address the flight conflicts through structured ascent or descent adjustments. 
In the high-speed corridors, a coarse-grained cylindrical aerial model is designed, consisting of four high-speed traffic loops and a central emergency corridor, to minimize the cross-corridor interference by isolating the high-speed traffic within dedicated airspace.
Moreover, a fine-grained horizontal grid model is designed, and the airspace is subdivided into dense 3D units to accommodate the complex structural and heavy population density of near-ground corridors. 
For the example in Fig. \ref{f2}, an aircraft firstly passes through three grids in a same corridor, then transfers to an adjacent corridor to avoid collisions with another aircraft, and finally to the third corridor to avoid collisions with ground obstacles.
During this process, the multi-modal wireless monitoring is necessary. 
Accordingly, these multi-functional corridors form a 3D air traffic network by optimizing corridor spacing and intersection design, which enables scaled and regularized operations of low-altitude aircrafts.

\begin{figure*}[!t]
    \centering
    \includegraphics[width=14.8cm]{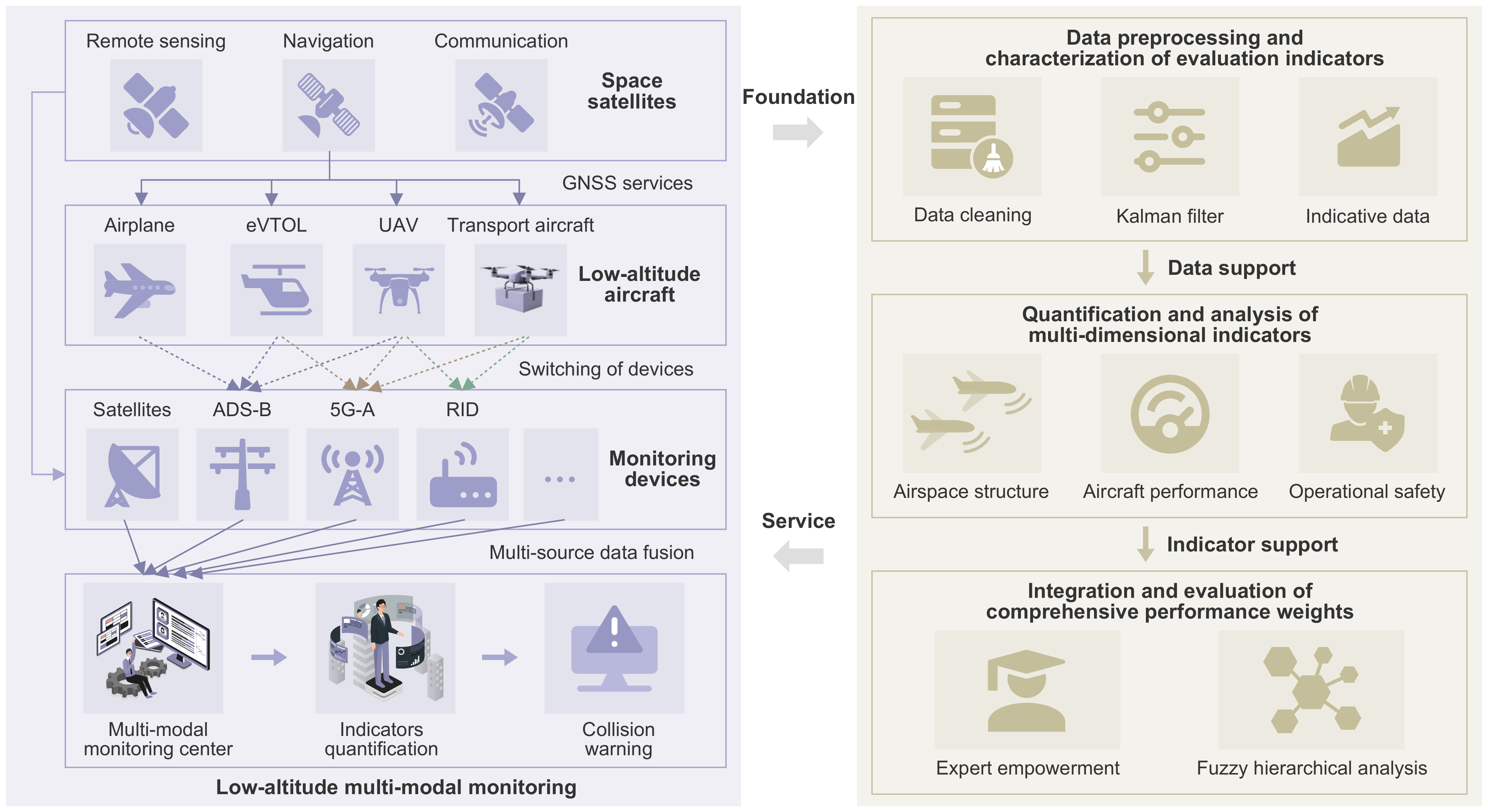}
    \caption{Collaborative monitoring for HLWN, supported by the low-altitude multi-modal  monitoring and multi-dimensional comprehensive evaluation.}
    \label{f3}
\end{figure*}

\subsection{Flight Rule for Corridor}

The rule design for corridors is to maximize the low-altitude airspace usability and minimize risks and costs.
In each corridor, there are entrance and output for aircrafts to enter or leave the corridor, and the traffic direction is one-way. 
If an aircraft want to change direction, it should exit the current corridor, and then enter an opposite direction corridor \cite{Sensors2021_corridor}.  
The velocity of each aircraft should be consistent in one corridor for the traffic smoothness \cite{ICC2024_traffic_UAS}.
Besides, the safe intervals should be satisfied in a corridor for safe spacing between aircrafts, and each small cube can accommodate only one aircraft at a particular instant. 
Both the velocity and safe interval can be monitored and guaranteed by the wireless technologies.

The intersections of the horizontal corridor and vertical corridor are inevitable, and the space is shared where intersect.
To coordinately manage the heterogeneous aircrafts, the 3D intersection in low-altitude airspace employs a crossed structure of coupled horizontal and vertical corridors, as shown in Fig. \ref{f2}. 
Through hierarchical safety intervals and adaptive aerial corridor conversion rules, such as the safe transition from one corridor to another with space restrictions, it can enhance the flight safety. 
In particular, the horizontal corridor dynamically adjusts lateral spacing based on the aircraft types  and employs multiple sensors and electronic fences to monitor the boundaries of corridors.
Besides, the buffer zones can be established to give priorities to the vertical traffic, and the aircraft entering these zones need to avoid possible collisions.
The aircraft with emergency mission can activate priority clearance rules, enabling coordinated optimization of 3D airspace utilization across multiple objectives.
Additionally, in high-density low-altitude environments, for safe and efficient  management of  low-altitude airspaces, the specific no-fly zones should be defined, covering areas such as military bases, high-speed railway stations, and civil aviation airports.

\subsection{Envelope Setting for Aircraft} 
The aircraft envelopes can facilitate the digital transformation and regularization of low-altitude traffic management by standardizing geometric shapes \cite{ICC_2024_CORRIDOR_Envelope}. 
Specifically, shapes such as spheres, ellipsoids, and cuboids define both the physical boundaries and mathematical models for computing collision risks. This design transforms the complex dynamic behaviors of aircraft into quantifiable geometric operations. 
Crucially, it also builds a fundamental technical defense line for the collision prevention through the precise definition of safety envelopes such as the radius of a sphere. 
Furthermore, it enhances the flight efficiency through streamlined shapes. 
The multi-modal monitoring is the basic for implementing the invisible electronic fences and envelopes.    
In addition, the dual-envelope structure comprises an inner physical envelope and an outer safety envelope, which embodies a core design philosophy for the low-altitude aircraft safety, shown in Fig. \ref{f2}. 
The main objective is to establish a progressive safety defense system to ensure efficient operations. The following analysis elaborates on this design.

\begin{itemize}    
    \item \textit{Physical Envelope}: 
The inner physical envelope precisely models the true shape of the aircraft, including protruding parts such as the airframe, rotors, and sensors.  
It directly addresses the core requirements for safety and operational functionality.  
Serving as the geometric reference for calculating physical collision risks, it provides critical supports for ensuring precise control of the aircraft in complex environments. 
Furthermore, the envelope provides the foundational spatial model for trajectory planning and obstacle avoidance algorithms.

\item \textit{Safety Envelope}: 
The outer envelope establishes a dynamic safety buffer around the inner physical envelope. 
This buffer creates a temporal collision warning window to trigger evasive operations, while accommodating system uncertainties such as localization errors, control latency, and environmental disturbances. 
Additionally, it ensures that flight operations meet required safety margins in line with the international collision avoidance standards.
\end{itemize}

In summary, the grid segment based  corridor design leverages extensive spatial coverage to  enable the  redundancy and adaptability. 
The model can also significantly help reduce delays and unnecessary circling caused by general conflicting trajectories.
The  envelope setting is significant for the safety flight and collision warning of low-altitude aircrafts, and also provides a basic model for the coordinated operation of diverse aircraft types in the high-density low-altitude airspace.

\begin{figure*}[!t]
    \centering
    \includegraphics[width=16cm]{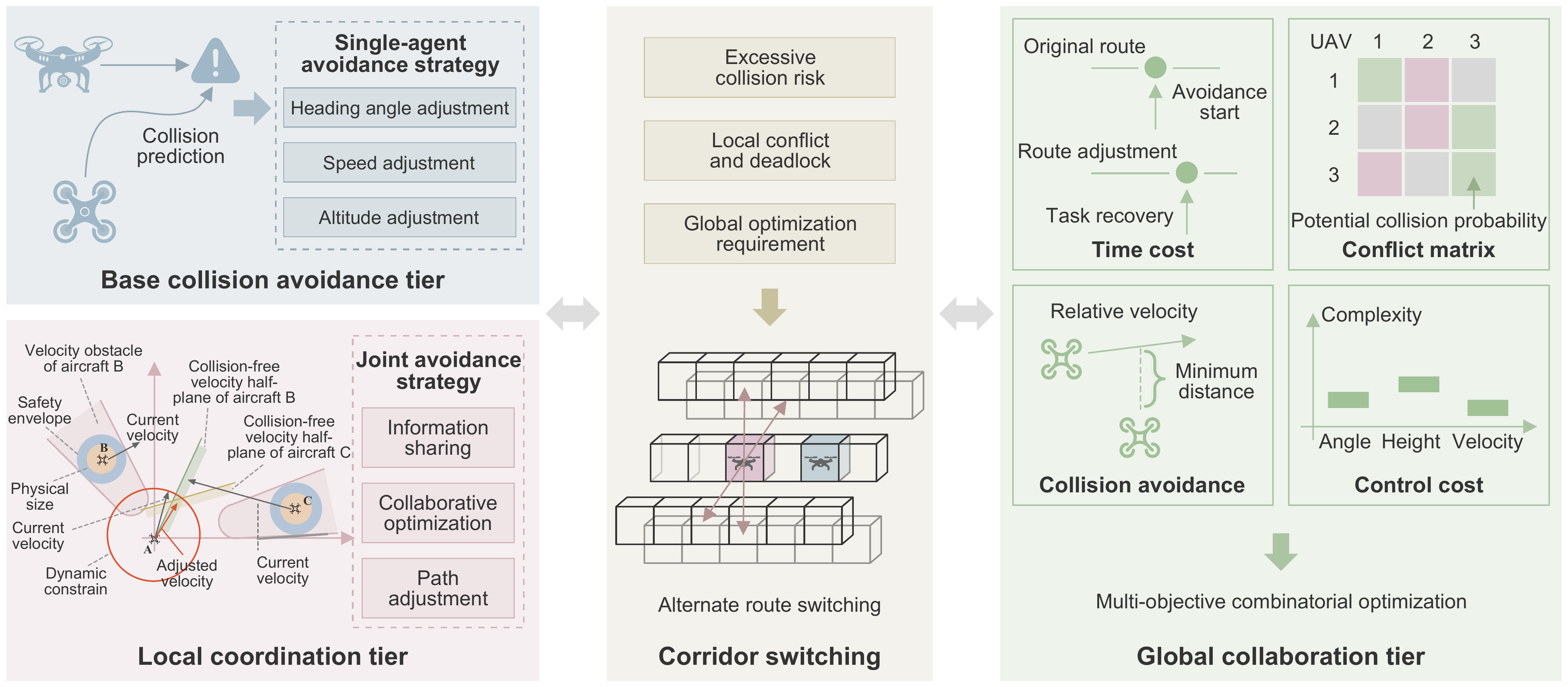}
    \caption{Collision avoidance based trajectory mechanisms for HLWN, including the  base collision avoidance, local coordination, corridor switching, and global collaboration.}
    \label{f4}
\end{figure*}

\section{Wireless Monitoring and Collision Avoidance}  

\subsection{Multi-Modal Monitoring}
The capacity constraints of airspace, communication and computation, limit the scale of HLWN, leading to three significant issues: conflicting demands for limited airspace resources, geographically imbalanced infrastructure coverage, and technical limitations in multi-source data integration \cite{sungeng_LAE_2024}.
For instance, the signals of  5G-A  may be obstructed by urban infrastructures, satellite communications suffer latency issues, and  the frequencies of RID experience serious interferences.  Therefore, the integration of navigation, communication and surveillance to achieve multi-source data fusion, is significant for  the multi-modal wireless monitoring of HLWN. 

As shown in Fig. \ref{f3}, the wireless monitoring  network is supported  by the 5G-A, ADS-B, RID, satellites, etc. 
The core lies in accommodating more communication modes to construct a real-time surveillance link between the low-altitude aircraft and multi-modal monitoring center. 
The  aircraft can select specific communication methods based on the environment by intelligent decisions, maximizing the performance of the HLWN, and thereby better adapting to complex and changeable application scenarios. 
In terms of the navigation subsystem, GNSS with the extensive coverage can help provide comprehensive monitoring \cite{zhangyifan_CJA} of large airspace areas. 
Besides, GNSS can also serve as auxiliary wireless communications for low-altitude aircrafts, ensuring that the aircraft can maintain contacts with the control center even in areas far from the ground monitoring center. 
In particular, the multi-modal monitoring can establish the anomaly detection and early warning mechanism \cite{yuanweijie_LAWN}. 
When anomalous conditions, such as irregular UAV trajectories, sudden speed fluctuations, or signal loss are detected, the system will promptly initiate a response.
For instance, if an aircraft enters the no-fly zone without permission, the monitoring system will respond promptly, giving operators valuable time to respond.

In terms of the multi-source data and multi-dimensional information from the multi-modal monitoring, the quantitative evaluation should be constructed  to provide reasonable   data support and evaluation. 
Firstly, the raw data from the multiple monitoring devices is received and preprocessed at the monitoring center by the data cleaning and Kalman filter to obtain the indicative data. 
Then, the correlations among various data sources are analyzed, and the high-level indicators such as the airspace structure, aircraft performance, and operational safety are constructed to enable the multi-dimensional assessment. 
Through the expert empowerment  and fuzzy hierarchical analysis, the integration and evaluation of comprehensive performance weights are obtained, as shown in Fig. \ref{f3}.
This process  provides a scientific evaluation and decision-making basis for the optimization of HLWN.

\subsection{Collision Avoidance Mechanism}

The collision avoidance based trajectory mechanisms for HLWN is shown in Fig. \ref{f4}, composed of 
the base collision avoidance tier, local coordination tier, corridor switching, and global collaboration tier, to ensure the safe operation of multiple low-altitude aircrafts within dynamic airspaces, detailed as follows.

\begin{itemize}
\item
The basic collision avoidance tier implements individual strategies for each aircraft through the  control mechanisms of heading angle adjustment, velocity modification, and altitude variation. 
The collision prediction can continuously analyze the sensor inputs for trajectory predictions.  
On this basis, this tier can autonomously initiate preventive operations at the single-agent level to reduce immediate collision risks.
\item
The local coordination tier works with the base tier to facilitate cooperative decision-making among adjacent UAVs through the information sharing and joint optimization. 
By implementing coordinated path adjustments, this tier resolves emergent conflicts by the isolated acting aircraft. 
To ensure the aerial safety, the tier also implements an expanded ring-shaped safety envelop that provides additional collision avoidance space. 
The collaborative mechanism employs consensus-based optimization algorithms to execute synchronized trajectory modifications, effectively addressing the localized conflicts.
\item
When a high  collision risk or potential deadlocks among multiple UAVs are detected by the multi-modal monitoring, the adaptive corridor switching rules are activated. 
It evaluates the global optimization requirements to identify the most efficient alternative corridors for all affected UAVs, to ensure the continued mission progressing while minimizing the disruptions.
\item
The global collaboration tier performs strategic route planning by the multi-objective optimization of collision-free trajectories to achieve the mission-level coordination. 
This tier should systematically balance the critical performance metrics including the time cost, collision probability, relative velocity, minimum separation distance, control cost, etc. 
By employing a conflict matrix that incorporates the parameters of angle, height, and velocity, the tier can evaluate operational complexity, assisting  in optimizing collision-free trajectories.
\end{itemize}

This collision avoidance mechanism achieves a dynamic balance between distributed decision-making at lower operational tiers and centralized optimization at higher strategic tiers. 
The balance also ensures the efficient and reliable collision avoidance in dynamic environments.

\subsection{Case Study} 
To validate the effectiveness of the proposed collision avoidance mechanism for HLWN, we evaluate the conflict resolution performance at a 3D corridor intersection, using MATLAB R2020b as the simulation platform for validation. 
Experiments are conducted in a $1,000\,\mathrm{m} \times 1,000\,\mathrm{m} \times 400\,\mathrm{m}$ cubic  simulated  space with the identical sets of the environmental factors. 
Three waypoints are set for each aircraft, namely the starting waypoint,  collision waypoint, and  ending waypoint, along with their corresponding arrival times, where the predefined trajectories of multiple aircrafts simultaneously converge at the collision waypoint of (500m, 500m, 300m) at the same arrival time, and the collision avoidance performance with envelop can be calculated. 
Furthermore, the parameter settings of multiple monitoring modes are as follows: the frequencies, transmission powers, and bandwidths of RID, 5G-A, ADS-B, and satellite (SAT) are set as (2.4e3, 3.5e3, 1.09e3, 1.6e3) MHz, (0.1, 4, 5, 3e3) W, (2, 100, 2, 8) MHz, respectively. 

\begin{figure}
    \centering
   \includegraphics[width=6.8cm]{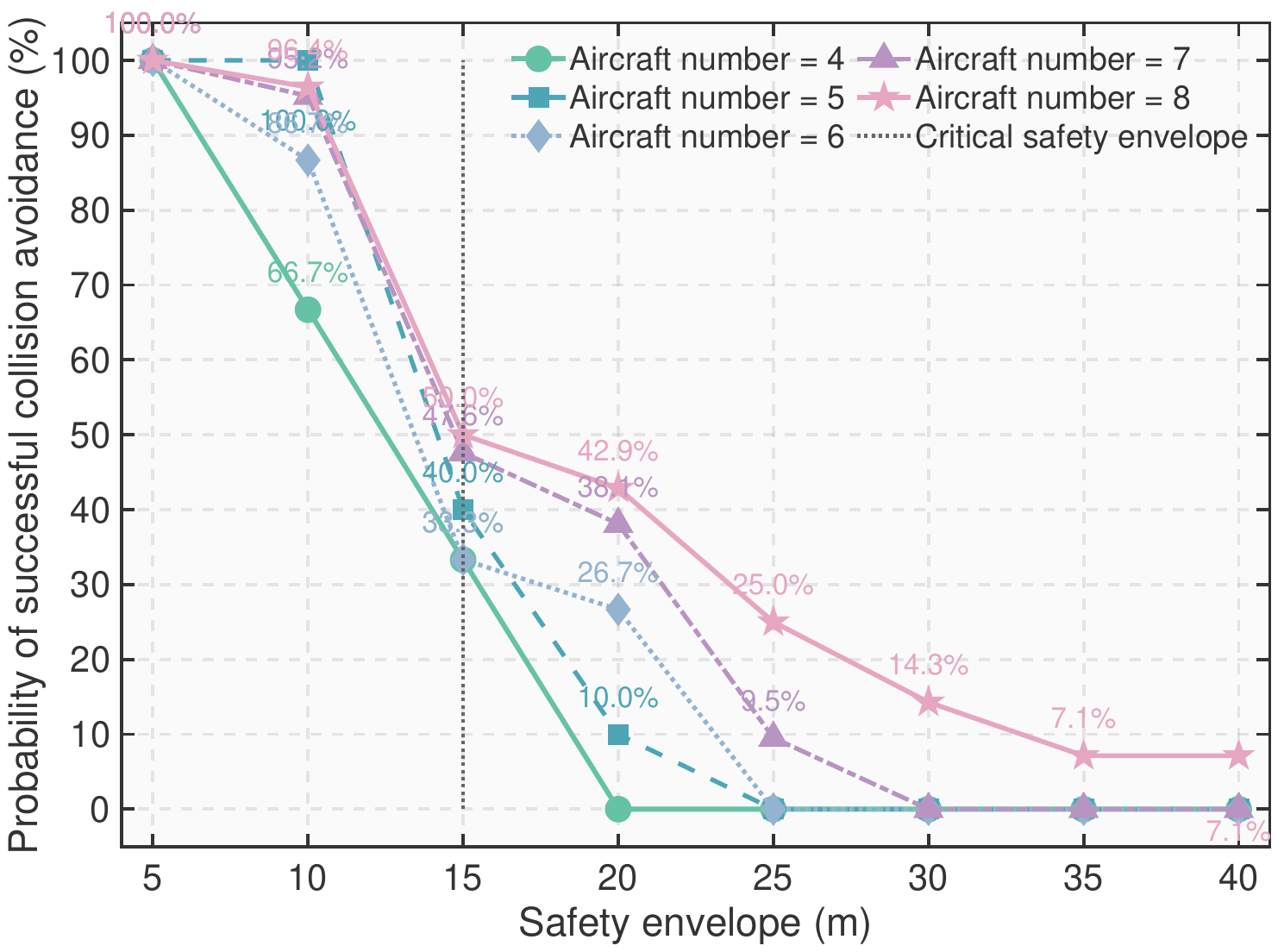}
   \vspace{-1mm}
    \caption{Performance of collision avoidance.}
    \label{fig5}
\end{figure}
\begin{figure}
    \centering
    \includegraphics[width=7cm]{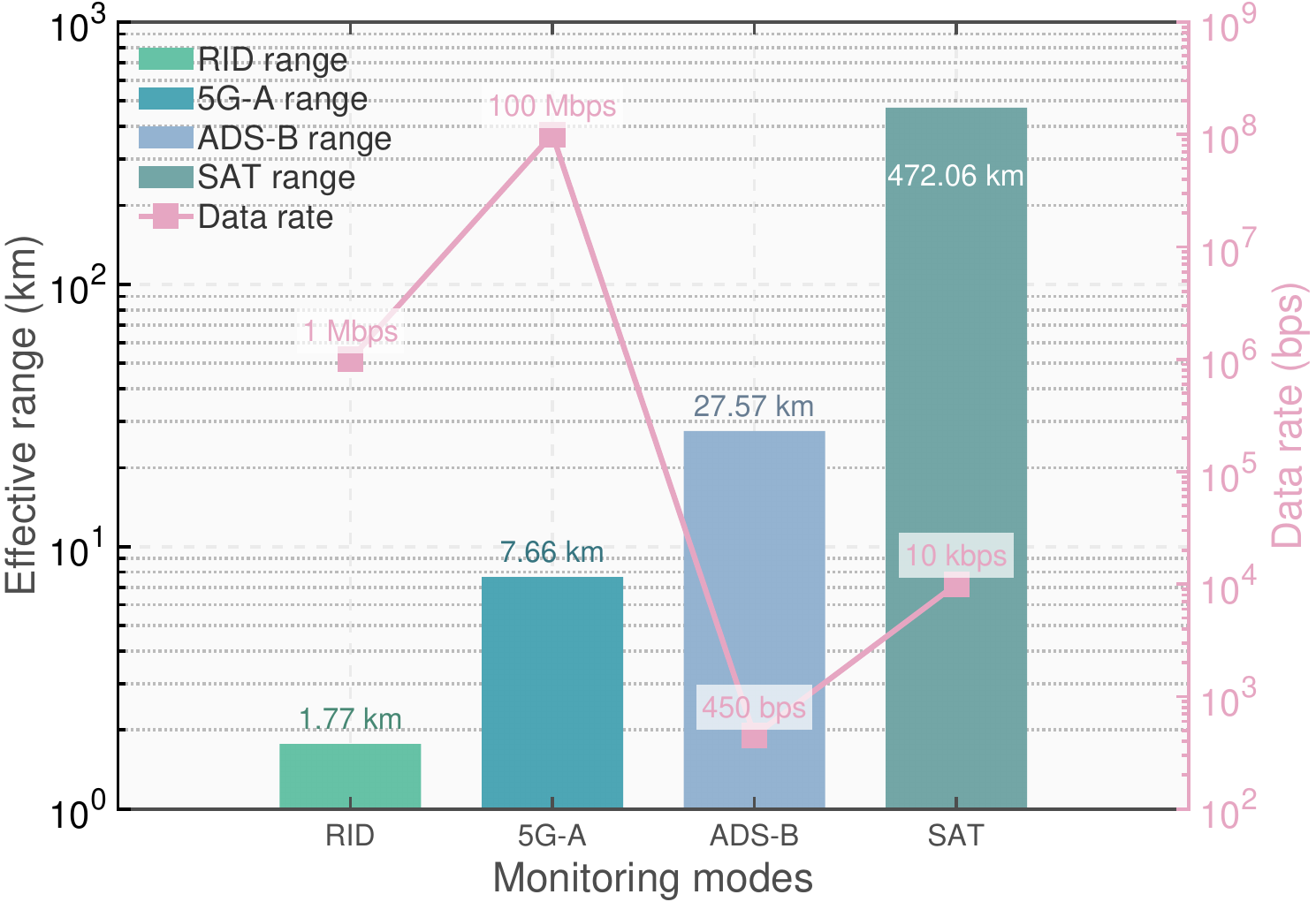}
       \vspace{-1mm}
    \caption{Effective range/data rate versus monitoring mode.}
    \label{fig6}
\end{figure}

We mainly analyze the effects of envelope and monitoring modes for HLWN.
Fig. \ref{fig5} illustrates the relationship between the collision avoidance probability and safety envelope for different aircraft densities. 
The probability of successful collision avoidance generally decreases for all  numbers of aircrafts with  the increasing size of the safety envelope, indicating the necessary to set the envelope.
Besides, there exists a tradeoff between the space efficiency and the size of safety envelope. 
Fig. \ref{fig6} compares the effective range and capable data rate of different wireless monitoring  modes, including RID, 5G-A, ADS-B, and SAT. 
It is noted that the 5G-A can provide high throughput and support data-intensive tasks.
Due to the large coverage, SAT can complement other modes to relay regional data.
Meanwhile, the ADS-B and RID can respectively provide the aircraft information of relatively  long and short ranges, enabling real-time airspace awareness. 
The integration of these wireless monitoring modes can enhance both the effective range and  data rate for the low-altitude air safety.

\section{Open Issues and Directions}  
The integration of HLWN into the low-altitude air traffic management shows great potentials to enhance the aircraft supervision  and  resource optimization  in dynamic  environments.
However, there still exist a series of challenges and research issues such as the  complexity of low-altitude environment, technological difficulties,  safety risks, etc.
Accordingly, we also analyze the possible directions to deal with these open issues for future researches.

\subsection{Open Issue}

\subsubsection{Low-altitude Environmental Complexity}
The growing density of low-altitude aerial operations from diverse aircrafts sharing airspaces brings key issues of dynamic airspace monitoring, real-time coordination, and multi-platform management. 
It requires developing an integrated intelligent sensing network combined with the adaptive coordination mechanism for heterogeneous aviation systems.

\subsubsection{Cooperation of Multi-modal Monitoring} 
In the scenarios with high-density air traffics, the conflict monitoring  face trade-offs between the computational efficiency and detection accuracy, while the multi-objective trajectory optimization struggles to balance multi-dimensional performances. 
It requires breakthroughs in the design of intelligent collision analysis algorithms, dynamic decision-making models, and heterogeneous sensor fusion technologies. 

\subsubsection{Security Risk}
The HLWN faces serious security risks, except for the previous safety of low-altitude airspaces, the data privacy for users, and the security for the wireless monitoring and transmission are also critical issues. 
Therefore, the stronger security system, better privacy protection method, and flexible certification processes are significant for HLWN.

\subsection{Future Direction}

\subsubsection{Digital Twin Empowered HLWN} 
As for the environmental complexity, integrating the digital twin technology can help construct the persistent virtual-to-real mapping for the low-altitude traffics, enabling high-quality predictions and emergency responses for HLWN. 
Besides, the consensus mechanisms can provide immutable audit trails for the aircraft trajectories, infrastructure states, and emergency protocols, ensuring regulatory compliance and operational transparency across the lifecycle of low-altitude traffics. 

\subsubsection{Adaptive and Intelligent Coordination} 
To deal with the integration of multi-modal monitoring, the adaptive coordination  to harmonize the UAV, eVTOL, and manned aircraft interoperability under constrained  low-altitude airspaces is significant. 
The intelligent multi-agent mechanism can be designed for the adaptive airspace partitioning and conflict resolution in mixed-operation environments. 
Particularly, the tradeoff of metrics for the multi-objective optimization also needs intelligent adaptive algorithm proposition. 

\subsubsection{Security Protection}
To address the security risks, the lightweight encryption and secure authentication can be deployed at the aircrafts to ensure the confidentiality.
Besides, the privacy-preserving mechanisms such as the differential privacy, can be applied to sensitive users.
The blockchain based smart contracts can also help autonomously validate the data provenance in distributed HLWN.


\section{CONCLUSION}

In this work, we have proposed HLWN for the low-altitude air traffic management, to achieve the safe heterogeneous  operation and optimal resource utilization. 
The grid based corridors,  flight rules, and safe envelopes have been designed for the safe trajectory of heterogeneous aircrafts.
We have  analyzed the collaborative wireless monitoring model based on the multiple space-air-ground surveillance facilities.
Then, we have developed the collision avoidance mechanisms, mainly including the adaptive trajectory planning and real-time hazard assessment for the safe corridor switching and trajectory.
Finally, we have discussed the open issues and future research directions.
This research on HLWN will improve the safety and efficiency for the low-altitude airspace.


\bibliographystyle{IEEEtran}
\bibliography{references}

\end{document}